\documentclass[twocolumn]{aastex701}
\makeatletter
\let\frontmatter@title@above=\relax
\makeatother

\usepackage{bm}

\usepackage{amsmath}
\usepackage{enumitem}

\newcommand{\funits}[1]{erg cm$^{-2}$ s$^{-1}$ \AA{}$^{-1}$}

\newcommand\target{AU~Mic}

\newcommand\jwst{\textit{JWST}}

\newcommand\kuit{\textit{Ku}}
\newcommand\kit{\textit{K}}

\received{July 9, 2026}
\revised{August 6, 2026}
\accepted{August 20, 2026}
\submitjournal{ApJ}

\shorttitle{Radio Micro-flare Signatures}
\shortauthors{Tristan et al.}

\begin{document}
\title{
Broadband 12--26 GHz Radio Radiation Reveals Evidence for Micro-flares on \target{}
}

\author[0000-0001-5974-4758,gname=Isaiah,sname=Tristan]{Isaiah I. Tristan}
\affiliation{Rice Space Institute, Rice University, Houston, TX 77005, USA}
\affiliation{Department of Physics and Astronomy, Rice University, Houston, TX 77005, USA}
\affiliation{Laboratory for Atmospheric and Space Physics, Boulder, CO 80303, USA}
\email[show]{isaiah.tristan@rice.edu}

\author[0000-0001-5643-8421,gname=Rachel,sname=Osten]{Rachel A. Osten}
\affiliation{Space Telescope Science Institute, Baltimore, MD 21218, USA}
\affiliation{Center for Astrophysical Sciences, Johns Hopkins University, Baltimore, MD 21218, USA}
\email{osten@stsci.edu}

\author[0000-0002-0412-0849,gname=Yuta,sname=Notsu]{Yuta Notsu}
\affiliation{Laboratory for Atmospheric and Space Physics, Boulder, CO 80303, USA}
\affiliation{Department of Astrophysical and Planetary Sciences, University of Colorado Boulder, CO 80305, USA}
\affil{National Solar Observatory, Boulder, CO 80303, USA}
\email{Yuta.Notsu@colorado.edu}

\author[0000-0002-9464-8101,gname=Adina,sname=Feinstein]{Adina D. Feinstein}
\affiliation{Department of Physics and Astronomy, Michigan State University, East Lansing, MI 48824, USA}
\email{adina@msu.edu}

\author[0000-0001-7458-1176,gname=Adam,sname=Kowalski]{Adam F. Kowalski}
\affiliation{Department of Astrophysical and Planetary Sciences, University of Colorado Boulder, CO 80305, USA}
\affiliation{Laboratory for Atmospheric and Space Physics, Boulder, CO 80303, USA}
\affil{National Solar Observatory, Boulder, CO 80303, USA}
\email{Adam.F.Kowalski@colorado.edu}

\correspondingauthor{Isaiah I. Tristan}

\begin{abstract}
We present sequential 12--18 and 18--26 GHz radio-band (\kuit{}, \kit{}) VLA observations of the 22 Myr dM1e star \target{}.
We detect two flares and two marginal events over a total of 3 contiguous hours on source, resulting in a radio flare rate of $\sim$1 flare hour$^{-1}$.
While this rate is consistent with previous \kuit{}-band observations, both flaring ($<$1 mJy) and quiescent ($\sim$0.4 mJy) flux densities are significantly lower. 
Furthermore, the quiescent spectral shape here is distinct, allowing for unique constraints on the radio-emitting sources of \target{}.
The time-averaged quiescent spectrum is best described by gyrosynchrotron radiation with a peak around 17~GHz and an optically thin spectral index of $\alpha\approx-0.6$. We estimate that the source regions have magnetic field strengths of $\sim$1~kG and cover a fraction of $<$0.5\% of the stellar surface, yet the instantaneous total electron kinetic energies are $\sim$10$^{28}$~erg. The power-law index describing the distribution of electrons with energy derived from the spectral index, $\delta \approx 2$, implies a near-continuous injection of electrons. This could arise from micro-flares that occur over the surface of \target{} that sustain the radio radiation.
One clear flare per band occurs, with decay-to-rise \textit{e}-folding time ratios of $3 - 4$, indicating magnetic trapping of the electrons. The \kuit{}-band flare is optically thick during the rise and peak times, indicating a peak frequency above 18 GHz.
Together, these quiescent and flaring characteristics suggest that continuous, unresolved micro-flaring and magnetic trapping dominate the non-thermal radio emission of active M-dwarf coronae.
\end{abstract}


\section{Introduction}
Understanding the physics of M-dwarf flares is crucial for evaluating the long-term habitability around these cool stars. Stellar flares on M dwarfs are much more energetic than their solar counterparts and involve a range of physical processes like particle acceleration, plasma heating, and mass motions \citep[see][for an overview]{Kowalski2024b}. The resulting high-energy radiation (particularly in the UV and X-ray regimes) may drive space weather, which can alter exoplanet atmospheres \citep[e.g.][]{chen21, louca23, nicholls23, doAmaral25}. However, the exact mechanism of particle acceleration in these extreme events is not well explained by solar models \citep{Kowalski2017A} and is an active topic of research \citep{Kowalski2022, Kowalski2024}.

Stellar flares emit radiation across the electromagnetic spectrum, but radio observations provide the most reliable trace of accelerated particles \citep{Osten2010} via gyrosynchrotron radiation or coherent emission \citep{Dulk1985}. Most time-resolved radio observations of stellar flares have historically been restricted to $\leq$10 GHz frequencies \citep{Gary1982, Gudel1989flare, Osten2002, Osten2005, Osten2006wideband, Villadsen2019, Bastian2022, Osten2026}, while observations at high frequencies (15--50 GHz) remain rare \citep{Fender2015, Plant2024, Tristan2025}.
Extremely high frequency observations (e.g., 230 GHz), often used to study debris disks, have also found short radio bursts that occur during stellar flares \citep{MacGregor2021}, necessitating an understanding of both stellar flares and disks for interpretation.
Filling the remaining 15--100 GHz gap is critical, as capturing the high-frequency, optically thin regime of the gyrosynchrotron spectrum allows us to directly estimate the energy distribution of the accelerated electron population \citep{Dulk1985}, which is an unconstrained parameter in stellar flare models \citep[e.g.,][]{Kowalski2025}.

Quiescent radio M-dwarf radiation is also generally attributed to gyrosynchrotron emission \citep[see the Güdel-Benz relation;][]{Gudel1993, Gudel2002}. 
However, there have been few direct spectroscopic confirmations and constraints with current-generation technology. 
For example, two studies of late-M dwarfs indicate gyrosynchrotron-based quiescence with potential turnover frequencies below 12 GHz \citep{Guirado2018, Plant2024}, but both rely on non-simultaneous wide-band data.
M-dwarf radio quiescence is highly variable \citep[see][and references within]{Tristan2026}, and the rate of change and the range of physical properties is unknown. Thus, more near-coincidence, wide-band observations are needed to improve constraints on the quiescent gyrosynchrotron signal, including how they influence emissions at other wavelengths and affect space weather.

Here, we present observations of \target{} (dM1e), an active flare star in the Beta Pic moving group \citep{Mamajek2014}. 
As one of the most X-ray luminous M-dwarfs within 10 pc \citep{Pallavicini1990, Leto2000} and a system that hosts a resolved debris disk \citep{Augereau2006, MacGregor2013, Grady2020} and transiting exoplanets \citep{Plavchan2020, Martioli2021}, \target{} is a valuable laboratory for studying both stellar flares and star-planet interactions.
The target frequently flares across the X-ray, UV, optical, and radio bands \citep{Robinson1993,Cully1994, Robinson2001, Redfield2002, Mitra2005, Hebb2007, MacGregor2020, Gilbert2022, Feinstein2022, Tristan2023, Notsu2025, Gibson2025}, including significant prior detections at 15 GHz \citep{Tristan2025}, making it an ideal target for follow-up high-frequency radio observations. 

For the first time, we take sequential Karl G. Jansky Very Large Array (VLA) observations of \target{} in the \kuit{} and \kit{} bands (Section~\ref{sec:data_reduction}) and identify a peak in the time-averaged quiescent spectrum (Section~\ref{sec:ana_quiet}).  The peak frequency is near 17 GHz, though the flux density is 2~--~8 times lower than previous \kuit{}-band observations with peak frequencies below 12 GHz.
The spectral shape and low polarization imply gyrosynchrotron radiation which arises from compact sources over a small fraction of the stellar disk, with magnetic field strengths around 1 kG.
We also analyze one flare per band captured during observations (Section~\ref{sec:ana_flare}), including significant detection of an optically thick spectrum during the rise and peak times.
We discuss our findings in the context of micro-flaring activity and stellar/solar flares in Section~\ref{sec:discussion}.
These data are collected as an ancillary project to \jwst{} GO 5311 (PI:~A.~Feinstein). 
Simultaneous data will be covered in other papers.

\section{Calibration and Data Reduction}
\label{sec:data_reduction}

The VLA provides 5 hours of \target{} observations on 2025 May 14 from approximately 09:50 to 14:50 UTC under program 24B-516 (PI:~I.~Tristan). The configuration is D-to-C with a maximum baseline of 3.4 km. Data are collected in the \kuit{}-band (12~--~18 GHz) and \kit{}-band (18~--~26 GHz) with $\sim$1.5 hours of on-source integration per band. The data are comprised of 48 and 64 subbands, respectively, with 128 channels of 1 MHz width per subband. The flux and phase calibrators are 3C286 and J2040-2507, respectively.

Initial calibration is performed using the standard CASA VLA Pipeline (v2024.1.1.22). We utilize the non-self-calibrated visibilities to avoid changing the variability of the target.
Following the pipeline, extensive manual inspection is performed. Frequency ranges heavily affected by satellite downlinks, bandpass edge effects, and a strong water vapor line (21~--~23 GHz) are excised. The data weights of the remaining visibilities are recalculated based on their empirical scatter using the CASA task \texttt{statwt}.

Initial imaging reveals \target{}, as well as three background point sources near the edge of the field of view (FOV). 
Primary beam corrections are applied, and integrated flux densities for the background sources (Table~\ref{tbl:sources}) are measured using \texttt{imfit} to account for unresolved extended emission.
To prevent FOV-edge phase-smearing from contaminating the target's light curve, we model and subtract the background sources directly from the visibilities using \texttt{tclean} and \texttt{uvsub}, and the phase center is shifted to \target{}.

\begin{deluxetable*}{lcccc}[!t]
\tabletypesize{}
\tablewidth{0pt}
\tablecaption{Radio Sources \label{tbl:sources}}
\tablehead{
\colhead{Source} & \colhead{RA} & \colhead{DEC} & \colhead{$\langle S_{Ku} \rangle$} & \colhead{$\langle S_K \rangle$}  \\ 
 & (ICRS) & (ICRS) & [mJy] & [mJy] 
}
\startdata 
\target & 20:45:10.10 $\pm$ 0.01 & --31.20.36.2 $\pm$ 0.05 & $0.363 \pm 0.003$ & $0.356 \pm 0.008$ \\
Field Source 1 & 20:45:02.16 $\pm$ 0.04 & --31.20.00.0 $\pm$ 0.11 & $0.39\pm0.02$ & $0.38 \pm 0.07$ \\
Field Source 2 & 20:45:01.50 $\pm$ 0.10 & --31.19.08.5 $\pm$ 0.30 & $3.08 \pm 0.24$ & \nodata \\
NVSS J204500-312121 & 20:44:59.97 $\pm$ 0.04 & --31.21.22.6 $\pm$ 0.11 & $1.16\pm0.08$ & \nodata \\
\enddata
\tablecomments{ 
The values for \target{} are estimated with visibility model fitting during times of quiescence, while all other values are the averages of synthesized images.
Both unnamed sources have faint detections in various surveys, but accurate cross-identification is beyond the scope of this work.
}
\end{deluxetable*}

\begin{figure}[t]
\centering
\includegraphics[width=1\linewidth]{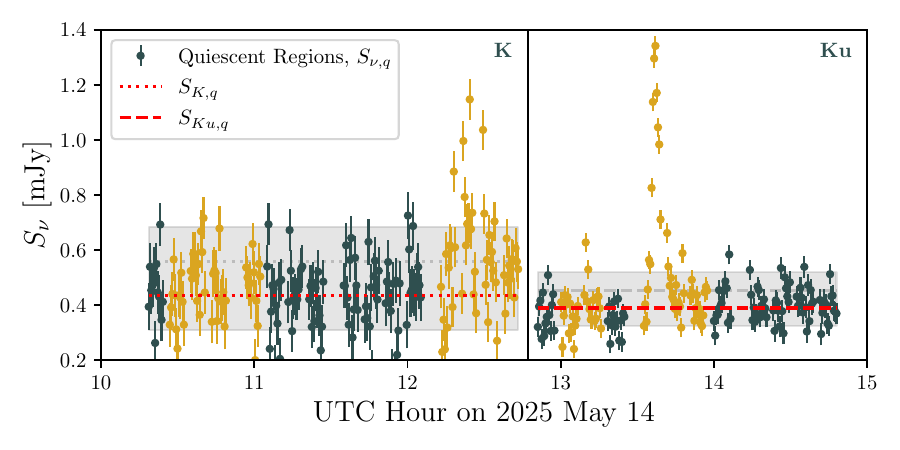}
\caption{\kit{}- (\textit{left}) and \kuit{}-band (\textit{right}) light curves with 30-second binning. The yellow circles indicate flaring or noisy times that were not used in fitting the quiescent. The red lines show the inverse variance weighted means (consistent with medians) of the quiescence, and the gray regions mark the $-1$$\sigma$ to $+2$$\sigma$ levels. 
\label{fig:light_curves}
}
\end{figure}

To generate light curves, the background source-subtracted visibilities are fit into 30-second temporal bins (Figure~\ref{fig:light_curves}) using \texttt{uvmodelfit}. 
Partial time bins at the end of scans are excised due to increased errors.
The \textit{RR} and \textit{LL} correlations are also fit separately in 1-minute bins and across quiescent times to calculate the Stokes \textit{I} and \textit{V} parameters,
\begin{align}
    I &= (RR + LL) / 2, \\
    V &= (RR - LL) / 2,
\end{align}
and the percentage of polarization is
\begin{align}
    \pi_c = V/I \times 100\%.
\end{align}

We isolate times devoid of flaring activity or increased noise and fit the visibilities in four equally spaced frequency ranges per band to extract the quiescent spectrum of \target{} (Figure~\ref{fig:spectra}).
We also perform this fit in 1-minute bins to create time-resolved spectra. 
Dynamic spectra are also created by averaging data over all baselines per subband in 1-minute integrations, but reveal no additional structures. 

\section{Quiescent Analysis and Results}
\label{sec:ana_quiet}

To assess the variability of \target{}, we test the light curves against a null hypothesis of constant quiescent flux density ($S_{\nu,q}$). The baseline $S_{\nu,q}$ is derived using an inverse variance weighted mean of the chosen times (Figure~\ref{fig:light_curves}). Further, a 5\% systematic uncertainty is added in quadrature to each time bin to account for any unforeseen systematics. We calculate a $\chi^2$ goodness-of-fit statistic, and the resulting probability of intrinsic variability, even in our chosen quiescent intervals, is $>$99.9\% for each band.

It is further well known that the $\geq$10 GHz radiation of \target{} varies drastically over time \citep{Cox1985, White1994, Leto2000, Tristan2025}.
Despite this intrinsic variability, \citet{Tristan2026} demonstrates that the shape of the radio quiescent spectrum remains similar on hour-to-day timescales, even with large $S_\nu$ variations. This stability implies similar radio-emitting sources.
We thus assume that the \kuit{}- and \kit{}-band quiescent spectra can be reasonably combined over 5 hours as a time-averaged representation of the low-activity state.

\begin{figure}[!ht]
\centering
\includegraphics[width=1\linewidth]{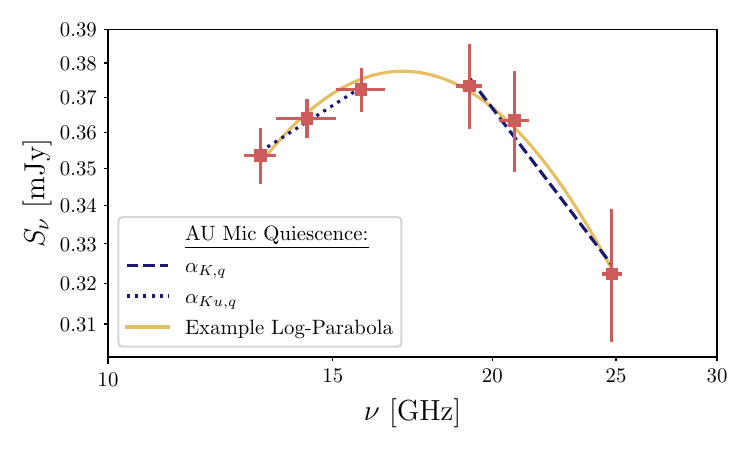}
\caption{\label{fig:spectra}
The quiescent spectrum of \target{} from non-simultaneous \kuit{}- and \kit{}-band data in log-log space. Dotted and dashed lines show the spectral index fits, respectively, and the solid line shows an example log-parabola fit. The horizontal bars display the frequency ranges covered. 
}
\end{figure}

The \kuit{}- and \kit{}-band exhibit rising and falling spectra, respectively. 
We employ Scipy's \texttt{curve fit} to calculate an inverse variance weighted linear fit in log space for the spectral index,
\begin{equation}
    \alpha = \frac{\log_{10}(S_{\nu_2}/S_{\nu_1})}{\log_{10}(\nu_2/\nu_1)}, \label{eq:specind}
\end{equation}
where $S_\nu$ is flux density and $\nu$ is the representative frequency with $\nu_1 < \nu_2$ (Figure~\ref{fig:spectra}). 
We derive the spectral peak using Monte Carlo (MC) simulations ($N=10^4$) to fit a log-parabola to the entire spectrum, also accounting for the systematic flux errors between bands \citep[5\% for \kuit{}, 7\% for \kit{};][]{Perley2017}. 
A log-parabola is chosen over a broken power-law as the majority of this frequency range appears to be in the transitional turnover region between optically thick and thin emission.
Derived properties are listed in Table~\ref{tbl:quiescence}.

\begin{deluxetable}{lll}[ht]
\tabletypesize{}
\tablewidth{0pt}
\tablecaption{Quiescent Properties \label{tbl:quiescence}}
\tablehead{
\textbf{Property} & \kuit{} & \kit{} 
}
\startdata 
\textbf{Per Band} & & \\ \hline
Central $\nu$ [GHz] & 15.0 & 22.0 \\
$S_{\nu, q}$ [mJy] & $0.39 \pm 0.07$ & $0.43 \pm 0.13$ \\
$\alpha_q$ & $+0.27 \pm 0.15$ & $-0.57 \pm 0.24$ \\ 
\hline \textbf{Modeled} & & \\ \hline
$\nu_{\text{peak}}$ [GHz] & $17.0^{+1.5}_{-1.9}$ & \\
$S_{\nu, \text{peak}}$ [mJy] & $0.38 \pm 0.02$ & \\
\enddata
\end{deluxetable}

Following \citet{Gudel2002}, the nonthermal electron number density follows a power-law in energy,
\begin{equation}
    n(E) = n_e (\delta - 1) E_0^{\delta - 1} E^{-\delta}~[\text{erg}^{-1}~\text{cm}^{-3}],
\end{equation}
where $E$ is kinetic energy, $E_0$ is the low-energy cutoff, $n_e$ is the total nonthermal electron number density above $E_0$, and $\delta$ is the electron power-law index.
We choose $E_0 = 10$ keV to stay consistent with the empirical equations of \citet{Dulk1985}, though the actual value may be much higher in M-dwarf flares \citep[see Section 3.5 of][for an extended discussion]{Tristan2025}.
Equation 39 of \citet{Dulk1985} can be used to estimate the magnetic field strength ($B$) of gyrosynchrotron source given the peak frequency ($\nu_{\text{peak}}$) and $\delta$, assuming a nonthermal electron column density ($n_e L$). 
The effective temperature ($T_{\text{eff}}$) of gyrosynchrotron radiation is then calculated using Equation 37 of \citet{Dulk1985}, where the emission is thought to come from the flaring layer where the source becomes optically thick \citep[i.e., at $\nu_{\text{peak}}$,][]{Osten2005}. 
Using $S_{\nu, \text{peak}}$ with the Rayleigh-Jeans Law and small angle approximation, we calculate a lower-limit of the brightness temperature at $\nu_{\text{peak}}$ over the disk of the source as
\begin{equation}
T_b = \frac{S_{\nu} c^2}{2 k_B \nu^2} \frac{d^2}{\pi R_*^2},
\end{equation}
where $c$ is the speed of light, $k_B$ is the Boltzmann constant, $d = 9.72~\text{pc}$ is the distance to the star \citep[][]{GaiaDR2}, and $R_* = 0.75~R_\sun{}$ is the stellar radius \citep[][]{Plavchan2020}.
Comparing these temperatures gives a filling factor of $f = T_b/T_{\text{eff}}$ for the source region emitting the gyrosynchrotron radiation. 
Multiplying this by the disk area of \target{} results in $A_{\text{source}} = f \pi R_*^2$. 
Using the same column density as before estimates a total number of $N_{tot} = n_e L \times A_{\text{source}}$ electrons. 
Following \citet{Smith2005}, the total instantaneous electron kinetic energy is 
\begin{equation}
    E_{\text{KE}} = (n_e V) \frac{\delta-2}{\delta-1}E_0= N_{tot}\frac{\delta-2}{\delta-1}E_0,
\end{equation}
where $V$ is the source volume.
If $\delta \leq 2$, this can be estimated instead by dividing the total energy integral by the total number integral,
\begin{align}
       \langle E_{\text{KE}} \rangle &= \frac{\int E \times n(E) dE}{\int n(E) dE} = \frac{\int^{E_{\text{max}}}_{E_{\text{0}}} E^{1 - \delta} dE}{\int^{E_{\text{max}}}_{E_{\text{0}}} E^{-\delta} dE}, \\
       E_{\text{KE}} &= N_{tot} \langle E_{\text{KE}} \rangle, 
\end{align}
with $E_{\text{max}} = 100$ MeV to set a sufficient high-energy cutoff.

We estimate $n_e L = 10^{15}~\text{to}~10^{16}$ cm$^{-2}$ based on source sizes and $n_e$ from optically thick flares on \target{} with a similar peaks of 16~--~18 GHz \citep[see Figure 10 of][]{Tristan2025}.
The optically thin spectral index (i.e., above the spectral peak) relates to $\delta$ by $\alpha\approx1.22-0.9\delta$. However, our calculated $\alpha_{K}$, which yields $\delta=1.99$, is likely diminished due to being near the peak frequency. A steeper slope calculated by the higher frequency points ($\approx$$-0.7$) or the high-end of the uncertainty range are more realistic values. Note that the empirical equations of \citet{Dulk1985} rely on an assumption of a homogeneous source with $2 \leq \delta \leq7$, with higher errors outside of this range. Despite these caveats, previous observations of flare peaks and quiescent gyrosynchrotron radiation from \target{} also exhibit $\delta \leq 2.1$ \citep[also see Section 3.1 of][]{Tristan2026}. Thus, we calculate results using a range of $n_eL$ and $\delta$ for comparison, including $\delta < 2$ (Table~\ref{tbl:modeled_variables}).
Note that $\delta$ does not have a dramatic effect on the results within the uncertainty range of $\alpha_K$.

\begin{deluxetable}{lll}[ht]
\tablecaption{Modeled Quiescent Source Results}
\tablehead{ \label{tbl:modeled_variables}
Variable & {$\delta=2.0^{+0.3}_{-0.3}$} & {$\delta=2.0^{+0.3}_{-0.3}$} \\
& $n_eL=10^{16}$ cm$^{-2}$ & $n_eL=10^{15}$ cm$^{-2}$
}
\startdata
\vspace{2pt} $B$ [G] & $679^{+84}_{-75}$ & $1526^{+124}_{-117}$ \\
\vspace{2pt} $T_{b}$ [K] & $4.5 \times 10^{6}$ & $4.5 \times 10^{6}$ \\
\vspace{2pt} $T_{\text{eff}}$ [K] & $2.2^{-0.4}_{+0.7} \times 10^{9}$ & $1.3^{-0.2}_{+0.4} \times 10^{9}$ \\
\vspace{2pt} $f$ [\%] & $0.20^{+0.05}_{-0.04}$ & $0.34^{+0.11}_{-0.07}$ \\
\vspace{2pt} $N_{tot}~[e^-]$  & $1.7^{+0.4}_{-0.4} \times 10^{35}$ & $2.9^{+0.8}_{-0.6} \times 10^{34}$ \\
\vspace{2pt} $E_{\text{KE}}$ [erg] & $3.1^{-2.6}_{+12.2} \times 10^{28}$ & $5.3^{-2.8}_{+21.0} \times 10^{27}$
\enddata
\end{deluxetable}

\section{Flare Analysis and Results}
\label{sec:ana_flare}

\begin{figure}[!ht]
\centering
\includegraphics[width=1\linewidth]{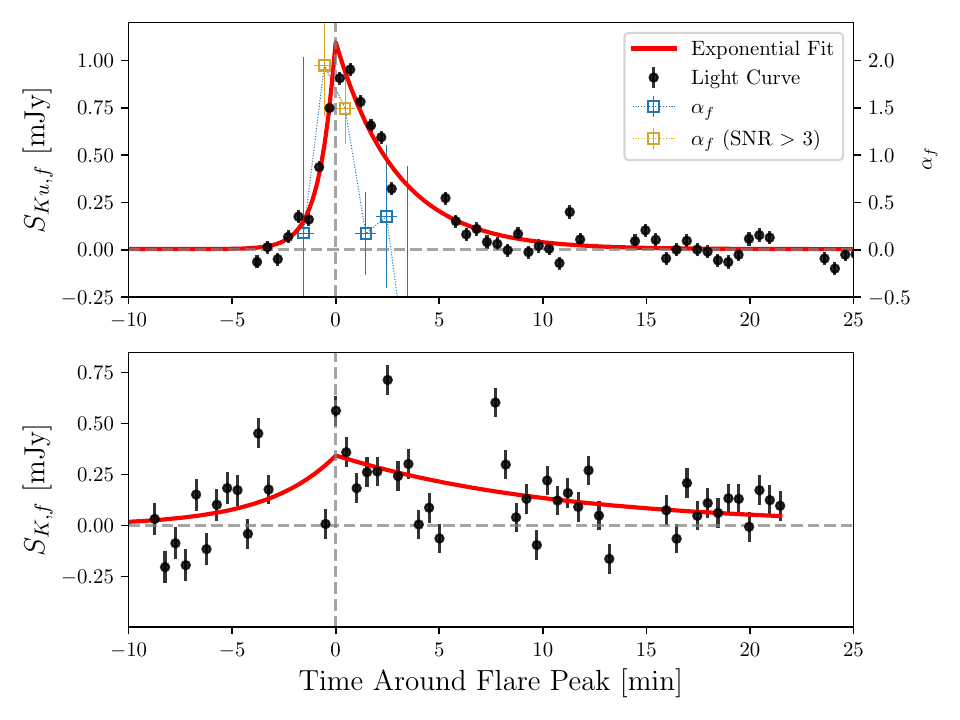}
\caption{\label{fig:flares}
Exponential fits to the \kuit- (\textit{top}) and \kit{}-band (\textit{bottom}) flares. The 1-minute binned flare-only spectral indices ($\alpha_f$) are also shown for the \kit{}-band flare. Derived parameters are listed in Table~\ref{tbl:flares}.
}
\end{figure}

Flares are selected by two consecutive points rising above  2$\sigma$ from the mean of the quiescent flux density, and this mean is subtracted before analysis. Only one clear flare per band is found in the 30-second light curve, with an additional marginal detection each.
Radiated flare energies are $E_{\text{Band}} = \mathcal{F} 4 \pi d^2 \Delta \nu$, where $\mathcal{F} = \int S_\nu dt$ is the fluence in units of erg cm$^{-2}$ Hz$^{-1}$ and $\Delta \nu$ is the bandwidth of the remaining frequencies.
Reported errors are the mean and standard deviation of an MC simulation which varies each time bin by its error and calculates $E_{\text{band}}$. 
We fit the flare profiles to rising and falling exponential models and calculate $e$-folding times for each ($\tau_{\text{rise}},~\tau_{\text{fall}}$), as well as the modeled peak flux density and time (Figure~\ref{fig:flares}).
We also calculate the impulsiveness index, $\mathcal{I} = I_{f,\text{peak}}/t_{1/2}$ \citep{Kowalski2013}, where $I_{f,\text{peak}}$ is the ratio of the flare-only flux density to the quiescent flux density and $t_{1/2}$ is the full-width at half-max of the light curve in minutes. 
We determine $t_{1/2}$ using the  exponential fit rather than data interpolation, due to a calibration gap in the \kuit{}-band flare and variability in the \kit{}-band flare.
Results are reported from the 30-second light curve analysis, and results using the 1-minute light curve are consistent within errors.

There are only two 1-minute spectra that achieve SNR$_\alpha$~$>$~3: the rise and peak of the \kuit{}-band flare (Figure \ref{fig:flares}). To determine the flare-only spectral index ($\alpha_f$), we subtract the quiescent spectrum from the 1-minute spectra, and this value falls slightly in the latter phase (Table~\ref{tbl:flares}).
Based on the 1-minute $RR$ and $LL$ light curves, we conservatively estimate that a $\pm$10\% polarization of the \kuit{}-band flare (also $\pm$30\% and $\pm$50\% of the \kuit{}- and \kuit{}-band quiescence, respectively) would be detected with an SNR~$>$~3. Finally, the disk-averaged $T_b$ remains below $10^8$~K for all flaring radiation.

\begin{deluxetable}{lll}[t]
\tabletypesize{}
\tablewidth{0pt}
\tablecaption{Flare Properties \label{tbl:flares}}
\tablehead{
\textbf{Property} & \kuit{} (2nd Flare) & \kit{} (1st Flare)
}
\startdata 
\textbf{Data} & & \\ \hline
$t_{tot}$ [min] & 12.10 & 29.20 \\
$E_{\text{Band}}$ [$10^{25}$erg] & $9.79\pm0.26$ & $7.94\pm0.74$ \\
$\alpha_{f,\text{rise}}$ & $1.94\pm0.53$ & \nodata \\
$\alpha_{f,\text{peak}}$ & $1.49\pm0.38$ & \nodata \\
$t_{\text{peak}}$ (UTC) & 13:37:06.0 & 12:24:24.0 \\ 
$S_{\nu,f,\text{peak}}$ [mJy] & $0.95 \pm 0.03$ & $0.71 \pm 0.07$ \\
\hline \textbf{Modeled} & & \\ \hline
$t_{\text{peak}}$ (UTC) & 13:36:23.2 $\pm$ 02.4 & 12:21:54.0 $\pm$ 30.6 \\
$S_{\nu,f,\text{peak}}$ [mJy] & $1.10\pm0.03$ & $0.34\pm0.03$ \\
$\tau_{\text{rise}}$ [min] & $0.77\pm0.06$ & $3.37\pm0.88$ \\
$\tau_{\text{decay}}$ [min] & $2.95\pm0.14$ & $10.67\pm1.96$ \\
$\tau_{\text{decay}}/\tau_{\text{rise}}$ & 3.82 & 3.16 \\
$\mathcal{I}$ [min$^{-1}$] & $1.09$ & $0.08$ \\
\enddata
\tablecomments{
$t_{tot}$ is the flare duration, $E_{\text{Band}}$ is the radiated energy within the band, $\alpha_{f}$ is the flare-only spectral index, $S_{\nu,f,\text{peak}}$ is the peak flare-only flux density, $\mathcal{I}$ is the impulsiveness index \citep{Kowalski2013}.
}
\end{deluxetable}

\subsection{A Short, Polarized Burst in the \kit{}-band Flare} 
\label{sec:burst}
\begin{figure}[!ht]
\centering
\includegraphics[width=1\linewidth]{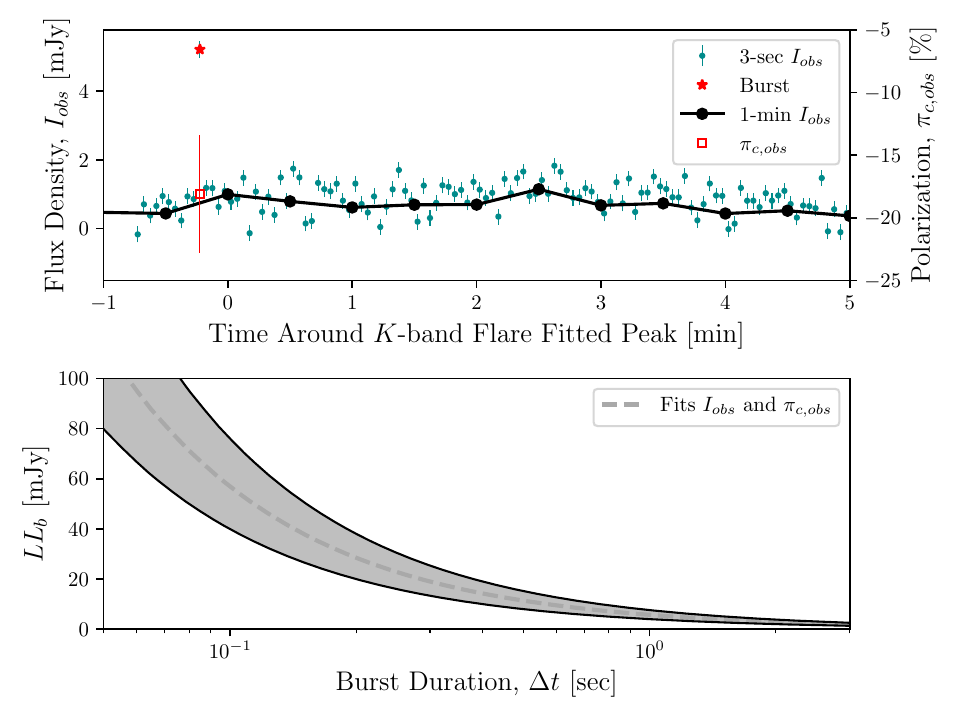}
\caption{\label{fig:burst}
\textit{Top:} Comparison of the 3-second and 30-second light curves. The 3-second burst properties are highlighted in red. Other $\pi_c$ are not shown due to high scatter.
\textit{Bottom:} Fitted $LL_b$ and $\Delta t$ that reproduce observed values. The gray shaded region covers 1$\sigma$ uncertainties.
}
\end{figure}

Millisecond solar radio bursts \citep[see][for an early review]{Benz1986} and fast, $<$2~GHz stellar radio bursts have been observed \citep{Osten2006wideband, Osten2008, Zhang2026}. However, sub-second to second-timescale high-frequency bursts are a relatively new topic, with only a few recent observations of these events at high frequencies \citep[e.g.,][]{MacGregor2021, Tristan2025}.
The morphology of the \kit{}-band flare suggests that it may be composed of several sub-bursts with durations of a few minutes or less. 
To investigate, we generate light curves at the VLA correlator dump rate of 3 seconds (Figure~\ref{fig:burst}). 
While some periods of the flare are reminiscent of minute-timescale bursts, their overall detection is not significant alone. 

The 3-second light curve does reveal a strong, slightly polarized radio burst within a single data bin, however. 
The burst spectrum is enhanced at all frequencies, indicating that it is not likely terrestrial radio frequency interference (RFI). 
Imaging this period reveals a strong detection near the source coordinates, while other 3-second \kit{}-band images exhibit marginal detections. We analyze the real and imaginary visibilities, finding no significant signs of RFI. The spectrum appears U-shaped, but this may come from the broad \kit{}-band water vapor line rather than a physical property.

Due to the time resolution, it is unclear what the true flux density, polarization, and physical mechanism are. 
One possibility is electron-cyclotron maser emission (ECME) based on solar millisecond busts \citep[][]{Fleishman1998}, which can be highly polarized ($\sim$100\%).
To estimate the flux density and duration from such emission, we recreate the observed 3-second integration using a 3-component model.
This model consists of the observed quiescent emission, an unpolarized flare with \textit{RR} and \textit{LL} flux densities equal to the observed \textit{RR}, and a 100\% \textit{LL} burst,
\begin{align}
    RR_{\text{model}} &= RR_{q, obs} + RR_{f,obs} \\
    LL_{\text{model}} &= LL_{q,obs} + RR_{f,obs} + LL_{b}\frac{\Delta t}{T_{int}},
\end{align}
where $q$ denotes quiescence, $f$ denotes the unpolarized flare, $b$ denotes the polarized burst, $\Delta t$ is the burst duration, and $T_{int} = 3$ seconds. The range of results are shown in Figure~\ref{fig:burst}.

For ECME emission, the frequency is tied to the electron cyclotron frequency, so the magnetic field can be estimated as
\begin{equation}
    B~[\text{G}] = \frac{\nu~[\text{Hz}]}{2.8\times10^6~s}, 
\end{equation}
where s is the harmonic number \citep{Dulk1985}. Across the \kit{}-band, the inferred magnetic field strength is 6.4~--~9.3 kG at the fundamental cyclotron frequency or 3.2~--~4.6 kG at the second harmonic. The broadband signal here could be explained by emission coming from a source region with a varying magnetic field strength. However, further details would require higher time resolutions to resolve.

\section{Discussion}
\label{sec:discussion}

\subsection{Quiescent Emission}
The potential causes of stellar radio radiation are many and varied, so we must discuss other possibilities that may produce similar signals.
The broadband spectral shape of the time-averaged quiescent spectrum ($\nu_{\text{peak}} \approx 17$ GHz, $\alpha \approx -0.6$) generally rules out gyroresonance emission, which would exhibit a much steeper cutoff and high polarization ($\pi_c>50$\%) after $\nu_{\text{peak}}$ \citep{Dulk1985}.
Thermal free-free (bremsstrahlung) emission from accelerated electrons in the stellar corona or wind is another possibility. 
While the associated spectrum is nearly flat at optically thin, high frequencies \citep{Dulk1985}, it could be argued that inhomogeneous, multi-thermal sources produce another spectral shape. 
However, optically thin free-free emission is estimated to be on the order of $S_\nu \approx 0.03$ for \target{} \citep{Leto2000}, and mass-loss rates calculated from values an order of magnitude higher that this and at 17 GHz are not physically sound \citep[see Section 3.3 of][]{Tristan2026}.
Another case could be electrons magnetically trapped in stable, closed magnetic loops, either in active regions or radiation belts similar to planetary magnetospheres.
This possibility would also specifically produce non-thermal gyrosynchrotron emission, though we may expect it at lower frequencies based on studies of ultracool dwarfs \citep[e.g.][]{Hallinan2015}. 
However, we would then expect the radio light curve to pulse or vary with the stellar rotational modulation \citep[as in early-type magnetic stars;][]{Leto2020}, which is not exhibited here nor in previous high-frequency observations of \target{}.
All three of these possibilities would be more stable and slowly varying with time, so the intrinsic variability of the light curve here also favors a more dynamic process, like micro-flaring.
Another process that may exhibit gyrosynchrotron radiation with intrinsic variability is particle acceleration due to turbulent waves in the corona \citep[see][]{Aschwanden2002}. However, the $\nu_{\text{peak}}$ would require high magnetic fields ($B>600$~G) and electron column densities ($n_eL > 10^{15}$ cm$^{-2}$) throughout the corona rather than in specific magnetic loops, so this is disfavored for this particular signal. A wave-driven signal is also likely to show short-timescale oscillations, which are not apparent here.
Finally, low polarization and $T_b$ values also rule out coherent emission in favor of gyrosynchrotron radiation, aside from the short 3-sec burst. With these considerations, we consider micro-flaring activity to be the most favored preferred scenario.

For comparison, more intense gyrosynchrotron flares are routinely around 20\% polarized outside of the peak phase \citep{Osten2005, Tristan2025}. Even lower polarization levels can be attributed to high optical depths or line-of-sight effects, namely where both sides of the flare loop are visible, resulting in an averaged-out polarization. Both are likely as $\nu_{\text{peak}}$ is observed and the loops cover $<$0.5\% of the surface area (Table~\ref{tbl:modeled_variables}). 
The power-law index inferred, $\delta \approx 2$, implies a hard distribution of electrons, which has been observed in the peaks of larger radio flares.
In this scenario, it better indicates a continuous injection of electrons, which would be indicative of unresolved micro-flaring events sustaining the radio radiation and heating the ambient plasma \citep[see Section 5 of][]{Gudel1993}.

Observations within recent years found the quiescent radio spectrum of \target{} to be dominated by a variable gyrosynchrotron source, with an additional underlying flat component of $S_{\kuit{}} = 0.64$~mJy \citep{Tristan2026}.
The quiescent flux density here is nearly half of just the flat component and appears wholly dominated by gyrosynchrotron radiation. 
Given the high level of radio variability seen in historical data \citep[e.g.,][]{Cox1985, Leto2000}, it is tempting to assume that this is a period of low magnetic activity.
However, $\nu_{\text{peak}} \approx 17$~GHz implies stronger source magnetic field strengths compared to the previously observed peak frequency $<$12 GHz. 
The quiescent $E_{\text{KE}}$ values are also fairly consistent with previous estimates \citep[see Figure 4 of][]{Tristan2026}, so it is unclear if the total energy budget and effects on space weather are consistent between the different states.

In relation to solar flares, estimates from \textit{RHESSI} solar X-ray data show median peak nonthermal powers of $10^{26}$~erg~s$^{-1}$ for solar flares at or below the \textit{GOES} C-class \citep{Hannah2008}.
\cite{Hannah2008} also reports that larger solar flares can exhibit peak nonthermal powers above $10^{27}$~erg~s$^{-1}$, which last longer (usually tens of minutes) than their smaller counterparts.
Thus, a superposition of M-dwarf micro-flares sustain nonthermal electron energies comparable to larger solar flares constantly, assuming that $E_{\text{KE}}$ remains consistent on second timescales.

As noted and referenced in Section~\ref{sec:ana_quiet}, there are several assumptions that are made when modeling the results listed in Table~\ref{tbl:modeled_variables}. The magnetic field strength ($B$) is particularly sensitive to the model used, compared to the other main variables like the filling factor ($f$) and non-thermal electron kinetic energy ($E_{\text{KE}}$). Regardless of $\delta$ and $n_eL$ values, $f<0.5$\%. While small, this factor is higher than the estimated flare footprint area of 0.01\% to 0.1\% of the stellar surface area \citep{Hawley1992, Hawley2003} and implies multiple micro-flaring sources. 
Note that this is somewhat higher than $f$ values used in stellar modeling of X-ray sources \citep[cf.~$f \approx 0.1$;][]{Takasao2020}, which can affect quiescent emission measure estimates \citep[see Section 5.2 of][]{Notsu2025}.
Similarly, $E_{\text{KE}}$ is centered within an order of magnitude of $10^{28}$ erg, which is generally high compared to solar flares.
However, $B$ changes drastically, in the context of a flaring loop, through the $n_eL$ range. This is because $B$ here is likely treated best as the average magnetic field strength between the various micro-flares at their optically thick-to-thin layers.
For more robust modeling, $B$ would change throughout the flare loop and the scale would likely be slightly different per source region, so these values presented should be considered tentative.

Micro-flares are generally thought to follow the same physical processes as larger flares \citep[e.g.,][]{Christe2008, Hannah2008}. Namely, a magnetic reconnection event along a magnetic loop accelerates electrons in the corona, which propagate through the loop and deposit energy into the chromosphere \citep{Fisher1985a,Fisher1985b,Kowalski2024b}. However, a single less-energetic event is unlikely to cause appreciable heating in the stellar atmosphere or produce an individual, detectable increase in ultraviolet, optical, or subsequent X-ray emissions.
These regimes also have other dominant quiescent emissions that make disentangling micro-flaring signatures difficult.
In contrast, the ambient free-free radiation at high radio frequencies is much quieter, which provides a unique opportunity to observe the action of accelerated particles from micro-flaring sources.
Further, this signal can be disentangled from distinct, strong events which clearly, but temporarily, dominate the radio regime.

\subsection{Flare Considerations}

The ratio of \textit{e}-folding times in both flares gives a profile that is consistent with average stellar flares \citep{Davenport2014}. 
In radio emission, the faster rise represents the impulsive accelerated particle injection while the decay is the gradual loss of magnetically trapped energetic electrons via Coulomb collisions and subsequent escape from the coronal loop \citep{Aschwanden1997, Bastian1998, Osten2026}.
Similar magnetic trapping during micro-flares could thus contribute to the sustained quiescent activity in active M-dwarfs.

The more intense \kuit{}-band flare is optically thick during the rise and peak, though the evolution beyond that is obscured (Figure~\ref{fig:flares}). This is consistent with many other small to moderate \kuit{}-band flares from \target{}, and the average radio $\nu_{\text{peak}}$ in M-dwarf flares is currently unknown. Note that there is no specific indication that the two flares are related to the micro-flares driving the quiescence, and similar $\nu_{\text{peak}}$ or $\alpha$ are not inherently expected.

\subsection{Non-interaction with the Debris Disk}

Finally, many radio M-dwarf flares exhibit small secondary peaks, which may be present in the \kuit{}-band flare here. To rule out echoing caused by the debris disk, we estimate a transmission rate (i.e., $e^{-\tau}$) of $>$99.99\% at a frequency of $\nu \leq 25$ GHz using
\begin{align}
    \tau &= \kappa_\nu\Sigma =\kappa_\nu/(\rho R_{\text{disk}}), \\
    \kappa_\nu &= 2.3(\nu/230~\text{GHz})^\beta~[\text{cm}^{2}~\text{g}^{-1}], \\ 
    \rho &= M_{\text{dust}}/(\pi R_{\text{disk}}^2 H), 
\end{align}
where $\tau$ is the optical depth, $\Sigma$ is the dust column density, $\kappa_\nu$ is the gas-poor dust absorption opacity with $1.7 < \beta < 2$ \citep{Beckwith1990}, $\rho$ is the mass density, $H$ is the scale height with an aspect ratio of $(H/R_{\text{disk}})\approx0.02$ \citep{Augereau2006}, $R_{\text{disk}}=40$~AU is the debris disk radius, and $M_{\text{dust}} = 7 \times 10^{25}$~g \citep{MacGregor2013}. Thus, we do not expect interaction with the debris disk to cause secondary peak echos, influence flare durations, nor affect spectral index or polarization measurements.

\section{Conclusions}

Using sequential VLA \kuit{}- and \kit{}-band observations of \target{}, we provide the widest high-frequency observations of \target{} to-date. 
We find that the quiescence is consistent with gyrosynchrotron radiation. 
However, the spectral shape is very different from previous observations.
This implies that the radiation comes from source regions of different properties despite their similar physical processes, which is consistent with the idea that the radio emission of M-dwarfs is driven by dynamic micro-flaring.
We also analyze two low-energy flaring events, which are consistent with standard radio flare findings, that occur during observations for completeness.
Our findings are summarized as the following.
\begin{enumerate}[noitemsep]
    \item The time-averaged quiescent spectrum resembles gyrosynchrotron radiation, with a peak of $\nu_{\text{peak}} \approx 17$ GHz and optically thin spectral index of $\alpha \approx -0.6$. This spectral index yields a power-law index of $\delta \approx 2$, which implies a hard energy distribution of electrons. Estimated magnetic field strengths are around $B=1$~kG, and this radiation is estimated to come from $<$1\% of the stellar disk. The instantaneous energies of the electrons are still high, at $E_{\text{KE}} \approx 10^{28}$~erg. These may be consistent with micro-flaring events.
    \item One significant flare is detected in each band, and both exhibit a $\tau_{\text{decay}}/\tau_{\text{rise}}$ ratio of $3 - 4$, consistent with the thought that electrons are magnetically trapped in the coronal loop.
    \item Neither the quiescence nor flares show significant amounts of circular polarization, where moderate levels would be detected at SNR~$>$ 3~levels. This is generally consistent with faint gyrosynchrotron radiation from stars. However, one sub-3-second burst exhibits a slightly polarized emission, indicating complex flaring processes that would require sub-second observing capabilities to resolve.
\end{enumerate}

Future work will include searching for other micro-flare signatures in the multiwavelength data from the campaign that would corroborate the findings here. 
For example, X-ray data can show similar stochastic variations. 
However, quiescent X-ray emission from high-temperature coronal plasma is spatially inconsistent with high-frequency radio sources \citep{Osten2005} and the variability is inconclusive with respect to rotational modulation \citep{Notsu2025}.
Alternatively, early studies of Far-UV spectra from \target{} attribute broad components of emission lines to micro-flares \citep[][]{Linsky1994, Pagano2000, Osten2006}.
Given that this campaign provides only a single point of reference, future follow-up multiwavelength observations at multiple epochs will likely be necessary to find correlations between the gyrosynchrotron spectrum and various micro-flaring signatures.
Future observations should also include higher wavelengths to continue filling in the $>$26 GHz coverage gap, as well as wider simultaneous frequency coverage to better constrain the shape of the radio spectrum.

\begin{acknowledgements}
I.I.T.~thanks Dr.~Zachary Berta-Thompson and Dr.~Andrea Isella for discussion on debris disks and stellar flares, Dr.~Steven Cranmer for discussion on time-steady coronal heating, Dr.~Girish Duvvuri and Dr.~Everett Schlawin for draft discussion, and Dr.~Heidi Medlin for VLA schedule coordination. We thank an anonymous reviewer for suggestions and discussion that led to the improvement of this manuscript.

This work is based in part on observations made with the NASA/ESA/CSA James Webb Space Telescope. The data were obtained from the Mikulski Archive for Space Telescopes at the Space Telescope Science Institute, which is operated by the Association of Universities for Research in Astronomy, Inc., under NASA contract NAS 5-03127 for JWST. These observations are associated with program \#5311.

VLA data is available at \url{https://data.nrao.edu}.
The National Radio Astronomy Observatory is a facility of the National Science Foundation operated under cooperative agreement by Associated Universities, Inc. 
\end{acknowledgements}

\facility{VLA}

\software{Astropy \citep{2022Astropy},
CASA \citep{CASA_2022},
Matplotlib \citep{Hunter2007},
Numpy \citep{harris2020array},
Scipy \citep{2020SciPy-NMeth},
CARTA \citep{carta2026}
}

\bibliography{bib}{}
\bibliographystyle{aasjournalv7}

\end{document}